\documentclass[aps,prb,twocolumn,superscriptaddress,floatfix]{revtex4}
\usepackage{graphicx}
\usepackage{dcolumn}
\usepackage{amsmath}
\usepackage{amssymb}
\usepackage{array}
\usepackage{calc}
\newcommand{\orto}{^{\circ}}

\begin{document}

\title{Temperature-induced Change in the Fermi Surface Topology in the Spin Density Wave Phase of Sr(Fe$_{1-x}$Co$_x$)$_2$As$_2$}
\author{Y.-X. Yang}
\affiliation{Laboratoire Mat\'eriaux et Ph\'enom$\grave{e}$nes Quantiques
(UMR 7162 CNRS), Universit\'e Paris Diderot-Paris 7, B\^at. Condorcet,
75205 Paris Cedex 13, France}
\affiliation{School of Physics and Technology, Wuhan University, Wuhan 430072, China}
\author{Y. Gallais}
\email{yann.gallais@univ-paris-diderot.fr}
\affiliation{Laboratoire Mat\'eriaux et Ph\'enom$\grave{e}$nes Quantiques
(UMR 7162 CNRS), Universit\'e Paris Diderot-Paris 7, B\^at. Condorcet,
75205 Paris Cedex 13, France}
\author{F. Rullier-Albenque}
\affiliation{CEA-Saclay, IRAMIS, Service de Physique de l'Etat Condens\'e
(SPEC URA CNRS 2464), F-91191 Gif-sur-Yvette, France}
\author{M.-A. M\'easson}
\affiliation{Laboratoire Mat\'eriaux et Ph\'enom$\grave{e}$nes Quantiques
(UMR 7162 CNRS), Universit\'e Paris Diderot-Paris 7, B\^at. Condorcet,
75205 Paris Cedex 13, France}
\author{M. Cazayous}
\affiliation{Laboratoire Mat\'eriaux et Ph\'enom$\grave{e}$nes Quantiques
(UMR 7162 CNRS), Universit\'e Paris Diderot-Paris 7, B\^at. Condorcet,
75205 Paris Cedex 13, France}
\author{A. Sacuto}
\affiliation{Laboratoire Mat\'eriaux et Ph\'enom$\grave{e}$nes Quantiques
(UMR 7162 CNRS), Universit\'e Paris Diderot-Paris 7, B\^at. Condorcet,
75205 Paris Cedex 13, France}
\author{J. Shi}
\affiliation{School of Physics and Technology, Wuhan University, Wuhan 430072, China}
\author{D. Colson}
 \affiliation{CEA-Saclay, IRAMIS, Service de Physique de l'Etat Condens\'e
(SPEC URA CNRS 2464), F-91191 Gif-sur-Yvette, France}
\author{A. Forget}
\affiliation{CEA-Saclay, IRAMIS, Service de Physique de l'Etat Condens\'e
(SPEC URA CNRS 2464), F-91191 Gif-sur-Yvette, France}

\begin{abstract}
We report electronic Raman scattering measurements of Sr(Fe$_{1-x}$Co$_x$)$_2$As$_2$ single crystals in their magnetic - Spin Density Wave (SDW) phase. The spectra display multiple, polarization-resolved SDW gaps as expected in a band-folding itinerant picture for a multiband system. The temperature dependence of the SDW gaps reveals an unusual evolution of the reconstructed electronic structure with at least one gap being activated only well below the magnetic SDW transition $T_N$. A comparison with temperature dependent Hall measurements allows us to assign this activated behavior to a change in the Fermi surface topology deep in the SDW phase, which we attribute to the disappearance of a hole-like Fermi pocket. Our results highlight the strong sensitivity of the low energy electronic structure to temperature in iron-arsenide superconductors. 

\end{abstract}

\maketitle

\section{INTRODUCTION}

The parent compound of iron-arsenide superconductors (Fe SC) are semi-metals whose electronic structure is composed of several Fermi surface sheets. Below $T_N$ they display a magnetic instability towards a metallic spin-density-wave (SDW) phase with a stripe-like antiferromagnetic structure \cite{Johnston}. Theoretical descriptions of the magnetic phase have taken both the itinerant and localized picture as a starting point \cite{Dai-Dagotto}. At low energy the magnetic instability can be described as a result of nesting between hole pockets centered at $\Gamma$ (0,0) and electron pockets centered at the $X (\pi,0)$ and $Y (0,\pi)$ points. The nesting wave-vector $Q_{AF}$=$(\pi,0)$ is consistent with the collinear spin structure observed in the magnetic phase \cite{Delacruz2008,Huang2008}. In this itinerant picture, the magnetically reconstructed Fermi surface results from band foldings at the nesting wave-vector $Q_{AF}$ and associated SDW gap openings at the new anti-crossing points between original and folded electronic bands in reciprocal space. Both band foldings and SDW gaps have been observed by optical spectroscopies \cite{Hu,Nakajima,Chauviere2011} and angle-resolved photoemission spectroscopy (ARPES) \cite{Yi2009,deJong,Kondo2010,Brouet2011}. However it has been argued that the complex reconstruction observed experimentally cannot be described by band folding alone, and may be accompanied by some degree of orbital order due to the associated $C_4$ symmetry breaking \cite{Yi,Brouet2011}. The importance of orbital degrees of freedom is linked to the on-site Hund's rule coupling \cite{Mazin2009,Kotliar-Hund, deMedici}, and can lead to significant departures from the simple itinerant nesting picture. It has been conjectured that Fe SC may be in an intermediate regime between itinerant and localized magnetism \cite{Mazin2009,Dai-Dagotto}.
\par
A consequence of the electronic reconstruction in the SDW phase is the strong reduction of the Fermi surface size as seen in quantum oscillations and Hall measurements on BaFe$_2$As$_2$ and SrFe$_2$As$_2$ single crystals \cite{Sebastian2008,Analytis2009, Sutherland, Rullier2009,Wen-Mazin-Hall}. The presence of a residual Fermi surface in the SDW phase can be attributed to the imperfect nesting between hole and electron pocket leading to an incomplete gapping of the Fermi surface and the presence of band edges lying close to the chemical potential. As a consequence the topology of the Fermi surface can be particularly sensitive to external perturbations such as doping, pressure or even temperature. Indeed strong effects of electron doping on the magnetically reconstructed Fermi surface, such as Lifshitz transitions, were reported in Co doped BaFe$_2$As$_2$  \cite{Liu,Mun-Canfield-TEP}, and could potentially control the onset of the superconducting phase in the underdoped region of the phase diagram. More recently, changes in the electronic structure with temperature were reported in ARPES experiments \cite{Dhaka,Brouet2013}. In these experiments, sizable band shifts were observed upon cooling in the paramagnetic phase, an unusual situation in ordinary metals with large Fermi surfaces. The shifts could be due to a strong sensitivity of the band structure to lattice thermal contraction. They could also originate from many-body effects which, in multiband systems, can lead to a significant Fermi surface reduction compared to DFT calculations, and render the low energy electronic structure sensitive to thermal effects even in the paramagnetic phase \cite{Benfatto}. The effect of temperature on the magnetically reconstructed electronic structure on the other hand, has received relatively little attention despite a presumably enhanced sensitivity due the reduced Fermi surface size as surmised early on by transport measurements \cite{Rullier2009}. Because of the potentially crucial role of the Fermi surface topology in setting both magnetic and superconducting orders, these temperature dependent band shifts can have a strong impact on the phase diagram of the Fe SC. They may also affect many properties such as transport and thermoelectric effects which are very sensitive to the carrier density and Fermi surface topology \cite{Rullier2009,Olariu2011,Mun-Canfield-TEP,Kuo-PRB2011}.

\par
Here we report temperature dependent electronic Raman scattering measurements in the SDW phase of pure and Co doped SrFe$_2$As$_2$ single crystals. The reconstruction of the electronic spectrum across the SDW transition at $T_N$ is consistent with multiple, polarization resolved, SDW gap openings as observed in previous spectroscopic studies. The overall energy scales of the SDW gaps decreases upon Co doping in agreement with the reduced $T_N$. However the evolution of the reconstructed electronic spectrum with temperature inside the magnetic phase of SrFe$_2$As$_2$ departs significantly from conventional mean-field behavior. In particular one of the SDW gap peak emerges only well below $T_N$, signaling an unusual temperature dependence of the SDW electronic structure. Comparison with temperature dependent Hall measurements allows us to assign this behavior to the disappearance of a hole band upon cooling below $T<$130~K where an optical transition, previously blocked by Pauli exclusion principle, becomes activated. Theoretical modeling of the Raman data in a simplified two-band model, with imperfect nesting and temperature dependent band shifts, is shown to account qualitatively for the observed activated behavior. Our results highlight the extreme sensitivity of the Fermi surface topology to temperature changes in the magnetic phase. They also show that previously observed band shifts in the paramagnetic phase extend deep in the magnetic phase with potentially more dramatic effects because of the reduced Fermi surface size. Finally they demonstrate the enhanced sensitivity of electronic Raman scattering to changes in the electronic structure in reconstructed phases like SDW.

\section{EXPERIMENTAL METHODS}

Single crystals of Sr(Fe$_{1-x}$Co$_x$)$_2$As$_2$ (Co-Sr122) were grown using a self-flux method. Single crystals of two different compositions were studied: $x$=0 and $x$=0.04. The parent single crystal SrFe$_2$As$_2$ was further annealed at 700$\orto$C for three weeks and then cooled down slowly to 300$\orto$C for several days. No annealing was performed on the $x$=0.04 crystal. The magnetic transition temperature, $T_{N}$, was determined both by transport and magnetization measurements yielding transition temperatures of 203~K and 137~K for $x$=0 and $x$=0.04 respectively, in good agreement with previous studies of the phase diagram of Co-Sr122 \cite{Canfield-Sr122, Gillett}. We note that, contrary to Co-Ba122 \cite{Ni,Chu-Ba122}, no clear splitting of the structural $T_S$ and magnetic $T_N$ transitions was resolved in Co-Sr122 \cite{Gillett, Canfield-Sr122}.

\par
Raman scattering experiments were performed on freshly cleaved single crystals. They were held in a vacuum of $\sim$ $10^{-6}~$mbar and cooled by a closed-cycle refrigerator. The spectra reported here were performed using the $\lambda = 532~$nm (2.33~eV) line of Diode Pumped Solid State (DPSS) laser and the $\lambda=488~$nm (2.54~eV) line of an Argon-Krypton laser. For both excitation energies an incident power of 10~mW was focused on an elliptical spot of dimension 50~$\mu$m$\times$120~$\mu$m, giving a power density of $\sim$ 220~W/cm$^2$.
 Reported temperatures take into account the estimated laser heating. It was first estimated by comparing the power and temperature dependences of the phonon frequencies. This estimate was then cross-checked by monitoring via a camera the onset of Rayleigh scattering by orthorhombic structural domains across the structural transition temperature as a function of laser power. Both methods yielded an estimated heating of 1$\:$K $\pm$ 0.2 per mW of incident laser power at 200~K. In terms of power density this translate into 4.5 10$^{-2}$$\:$K $\pm$ 10$^{-2}$  per W/cm$^2$.  At 100~K the former method gave a similar estimate of 1$\:$K $\pm$ 0.5 per mW.  
\par
For the high energy Raman spectra (150 to 5000 ~cm$^{-1}$), the scattered light was analyzed by a single stage spectrometer equipped with a 600~lines/mm grating and high-pass edge filters (RazorEdge, Semrock) to block the elastically scattered light. For the low energy spectra ($\leq$ 150 ~cm$^{-1}$) a triple stage spectrometer (JY-T64000) equipped with 1800~lines/mm gratings was used. In both set-ups the light was detected by a liquid nitrogen cooled CCD detector. Raman spectra were all corrected for the corresponding instrumental spectral response. The imaginary part of the Raman response was then extracted using $\chi''\sim(1+n(\omega,T))^{-1} \times I$, where $n(\omega,T)$ is the Bose-Einstein distribution function, $\chi''$ the imaginary part of the Raman response and $I$ the measured Raman intensity corrected for the instrumental response. 
\par
\begin{figure}
\centering \includegraphics[clip,width=0.9\linewidth]{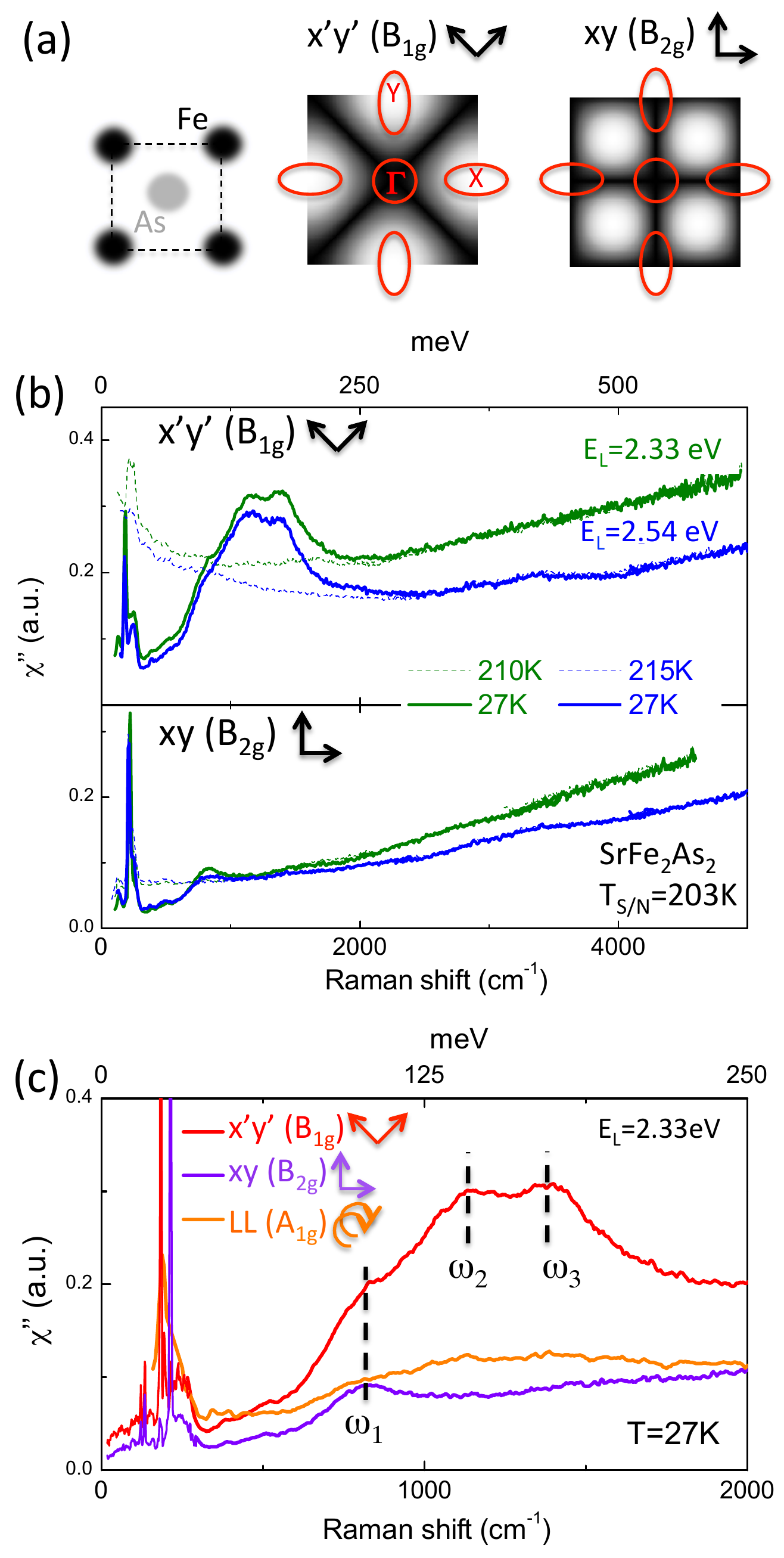}
\caption{(a) FeAs square plane and k-space structure of the Raman form factors or vertex in $x'y'$ ($B_{1g}$) and $xy$ ($B_{2g}$) symmetries (see text). 1 Fe unit cell has been used. $x$ and $y$ refer to the directions of the Fe-Fe bonds while $x'$ and $y'$ are oriented along Fe-As bonds. For $x'y'$ and $xy$ symmetries, the location of the Fermi pockets have been superimposed in red. (b) Wide energy Raman spectra in $x'y'$ ($B_{1g}$) and $xy$ ($B_{2g}$) symmetries for SrFe$_2$As$_2$ ($T_{S/N}$=203~K).  (c) Polarization resolved spectra at low energy and at 27~K. }
\label{fig1} 
\end{figure}

Polarization resolved spectra were obtained by using different incoming and outgoing photon polarizations, and by aligning them with the crystallographic axis in-situ via a piezo-rotating stage. The symmetries mentioned in the paper all refer to the tetragonal crystal structure with 1 Fe unit cell and are thus rotated by 45 degrees with respect to the 2 Fe unit cell which takes into account the alternating As atoms. We will mostly focus on two complementary symmetries (referred to as $B_{1g}$ and $B_{2g}$ symmetries in the tetragonal 1 Fe unit cell) which correspond to the $x'y'$ and $xy$ photon polarizations configurations (see Fig. \ref{fig1}(a)). 
\par
For a single orbital tight-binding model on a square lattice, the Raman form factors (or vertex) have simple $\cos k_x\cos k_y$ and $\sin k_x \sin k_y$ dependences for $x'y'$ ($B_{1g}$) and $xy$ ($B_{2g}$) symmetries respectively (see figure \ref{fig1}(a)). Given the Fermi surface topology of the iron-pnicitides, these $k$ dependences imply that the $x'y'$ ($B_{1g}$) symmetry samples mostly the electron pockets at $X$ and $Y$ points, while the $xy$ ($B_{2g}$) symmetry does not coupled significantly to neither electron nor hole pockets (see Fig. \ref{fig1}(a)). In a multi-orbital system like the iron-pnictides, these simple $k$ dependences are valid only in the case of intra-orbital excitations \cite{Belen2013}. They may nevertheless be qualitatively valid at low energies where intra-orbital excitations dominate the spectra. We also note that Raman scattering data in the superconducting state of Ba(Fe$_{1-x}$Co$_x$)$_2$As$_2$, along with calculations of the Raman form factors using the full Density Functional Theory band structure, indicate that the $x'y'$ ($B_{1g}$) symmetry probes mostly the electron pockets, as expected from the simple $k$ dependences described above \cite{Muschler,Chauviere2010, Mazin}.

\section{RAMAN RESULTS}

Raman spectra up to 5000 ~cm$^{-1}$ for $x$=0 are shown in Fig. \ref{fig1}(b) for the two different polarization configurations both slightly above and well below $T_{N}$=203~K. The directions of the incoming and outgoing polarizations with respect to the FeAs square plane are depicted in the Fig. \ref{fig1}(a). The spectra are shown for both 2.33~eV (in blue)  and 2.54~eV (in green) excitation energies. They exhibit a continuum that extends up to the highest energy measured. At low energy ($\leq$ 300~cm$^{-1}$) the continuum is superimposed by sharp peaks due to Raman active optical phonons whose behavior will be described elsewhere. The continuum is mostly of electronic origin and displays a pronounced reconstruction upon entering the SDW phase for $\omega\leq$ 2000~cm$^{-1}$. This is the case both for $x'y'$ ($B_{1g}$) and $xy$ ($B_{2g}$) symmetries. Similar reconstruction is seen when exciting with 2.33~eV and 2.54~eV laser lines. The reconstruction is ascribed to $Q=(\pi,0)$ band folding induced gap openings upon entering the SDW phase. The gap openings are evidenced by a strong suppression of the low energy excitations below $\omega\sim$700~cm$^{-1}$ and the simultaneous activation of symmetry dependent optical transitions across the gaps, hereafter referred to as SDW gap peaks, in the 700-2000~cm$^{-1}$ energy range.
\par
The temperature dependent electronic continuum is superimposed on a featureless and raising background which is essentially temperature independent above 2000~cm$^{-1}$ and extends well beyond 5000~cm$^{-1}$. Its overall magnitude and slope depend on the excitation energy used, hinting that part of it at least may be due to luminescence either from recombination processes in the bulk or from residual surface adsorbates. We note that in contrast to an earlier Raman study in BaFe$_2$As$_2$ (Ba122) \cite{Sugai}, we did not observed in Sr122, nor in Ba122, any evidence for two-magnon excitations which in a local moment picture are expected to arise in the magnetic phase at energies above 2000~cm$^{-1}$ \cite{Sugai,2magnon-Devereaux}. This absence is in stark contrast to insulating antiferromagnetic cuprates where two-magnons excitations dominate the Raman spectra \cite{Lyons-magnon, Sugai-magnon}. It appears that, at least from the Raman scattering perspective, the magnetism in 122 iron-pnictides is better described by a itinerant SDW-like rather than a local moment point of view.

\begin{figure}
\centering \includegraphics[clip,width=0.85\linewidth]{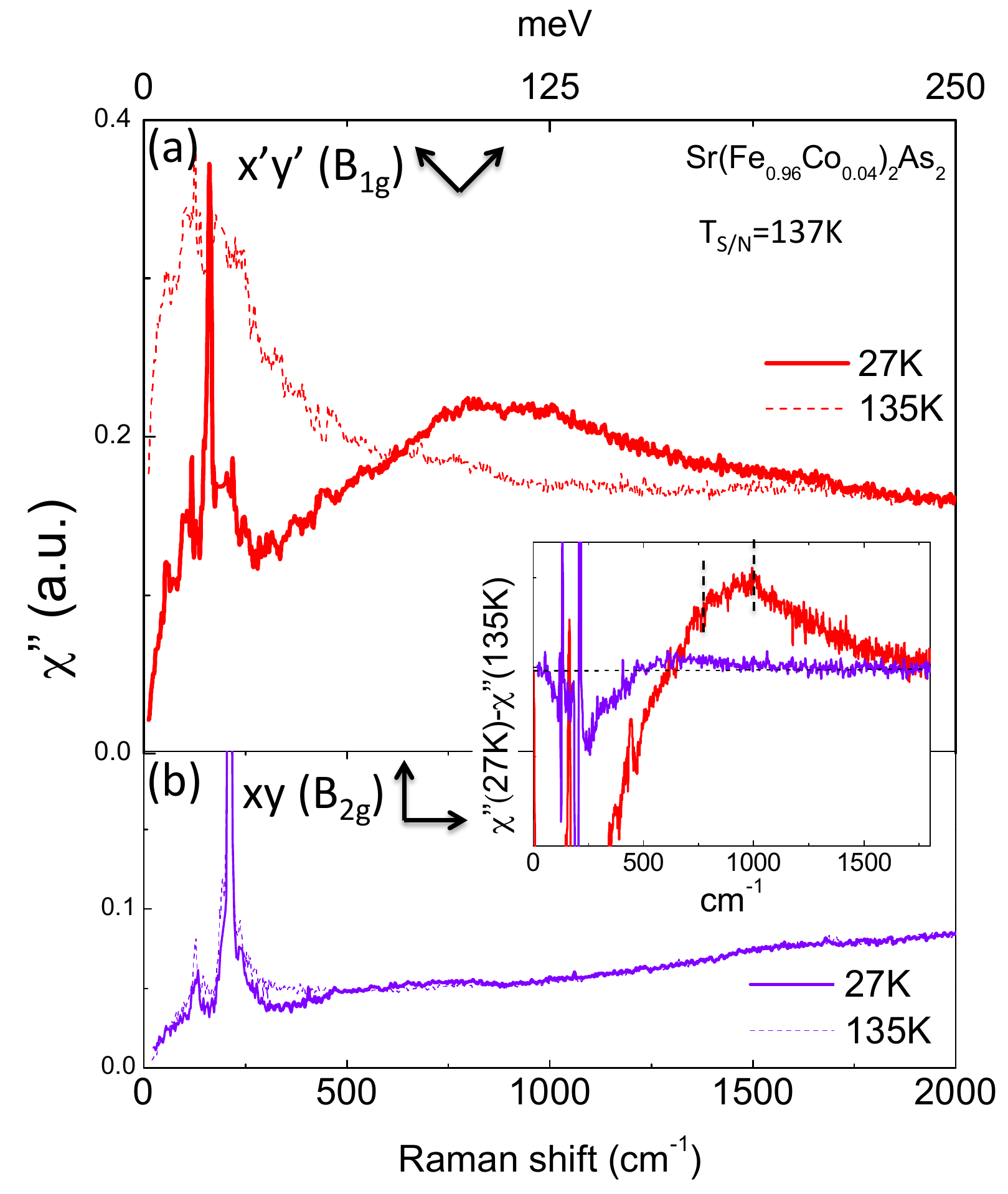}
\caption{ Raman spectra in $x'y'$ ($B_{1g}$) and $xy$ ($B_{2g}$) symmetries for Sr(Fe$_{1-x}$Co$_x$)$_2$As$_2$ ($T_{N}$=137~K) using 2.33~eV excitation energy. The inset shows the subtraction between the response at low temperature (27~K) and the response at $\sim$ $T_{N}$ for both symmetries. The dotted lines highlight the positions of the two kinks observed at $\sim$ 750 and 1000~cm$^{-1}$ in the $x'y'$ spectrum.}
\label{fig2} 
\end{figure}

\par
We now focus on the low energy part of the spectrum ($\leq$2000~cm$^{-1}$) shown in Fig. \ref{fig1}(c). In addition to the $x'y'$ and $xy$ linear polarization configurations, circular polarized incoming and outgoing photons (LL) are used to extract the $A_{1g}$ symmetry. At low temperature, at least three symmetry dependent SDW gap peaks are observed and are indicated by dash lines. The lowest energy peak, labelled $\omega_1$, is observed in both $x'y'$ ($B_{1g}$) and $xy$ ($B_{2g}$) symmetries at 820~cm$^{-1}$. The two peaks at higher energies, $\omega_2$ and $\omega_3$, are clearly resolved in $x'y'$ symmetry only, and are located at 1140~cm$^{-1}$ and 1420~cm$^{1}$ respectively. Only a weak and broad structure, with no clearly resolved peaks, is observed in LL ($A_{1g}$) symmetry. The overall symmetry dependence and the energy scales of the high energy SDW gap peaks are similar to previous results on Ba122 when taking into account the higher $T_N$ \cite{Chauviere2011}. Probably due to the higher $T_N$, the Raman spectra in Sr122 show both a stronger reconstruction and more SDW gap peaks than in Ba122. For example, in Ba122 ($T_N$=138~K) the low temperature $x'y'$ spectrum is dominated by a single peak at 900~cm$^{-1}$ or 9.5~$k_BT_N$, while in Sr122 two peaks, located at 8.1 and 10.1~$k_BT_N$ respectively, can be clearly resolved in the same symmetry. The energy scales of the Raman SDW gaps are also close to the one extracted from optical conductivity measurements in both systems \cite{Hu,Hancock}. Interestingly, in optical conductivity measurements the single peak observed in twinned Ba122 and Sr122 crystals at around 10~$k_BT_N$ was shown to consist of two peaks, resolved in $\sigma_{xx}$ and $\sigma_{yy}$ respectively, on detwinned single crystals \cite{Nakajima-PNAS}. In the case of Sr122, their energies are close to the energies of the $\omega_2$ and $\omega_3$ peaks observed in the $x'y'$ Raman spectrum.
\par
Theoretical calculations of the Raman and optical conductivity spectra within a five band model suggest that the $\omega_2$ and $\omega_3$ peaks are due to SDW gap openings in different locations in momentum space, both sampled in $x'y'$ symmetry: at the electron pockets around the $X$ and $Y$ points of the Brillouin zone respectively \cite{Belen2013,Sugimoto}. The low energy $\omega_1$ peak on the other hand is not observed in optical conductivity \cite{Hu,Hancock}. Its exact assignment is uncertain, but it may result from an optical transition between folded bands away from the $\Gamma- X$ and  $\Gamma-Y$ directions which are sampled in $xy$ symmetry \cite{Belen2013}.
\par
A similar but somewhat weaker reconstruction is observed for $x$=0.04 ($T_{N}$=137~K) as shown in Fig. \ref{fig2}. As in the undoped case, the strongest reconstruction is observed in the $x'y'$ spectra where a broad structure is observed between 600~cm$^{-1}$ and 1700~cm$^{-1}$ in the SDW state. By contrast only a relatively weak suppression below 450~cm$^{-1}$ is observed in the $xy$ spectra. Subtraction of the $x'y'$ spectrum just above $T_{N}$ allows us to identify at least two kinks within the broad structure, at 750 (8 $k_BT_N$) and 1000~cm$^{-1}$ (10.6 $k_BT_N$) respectively, close to the energy scales of the $\omega_2$ and $\omega_3$ peaks in undoped Sr122. This suggests an overall scaling of the SDW gap energies with $T_N$ upon Co doping at least up to $x_{Co}$=0.04.

\begin{figure}
\centering \includegraphics[clip=,width=0.85\linewidth]{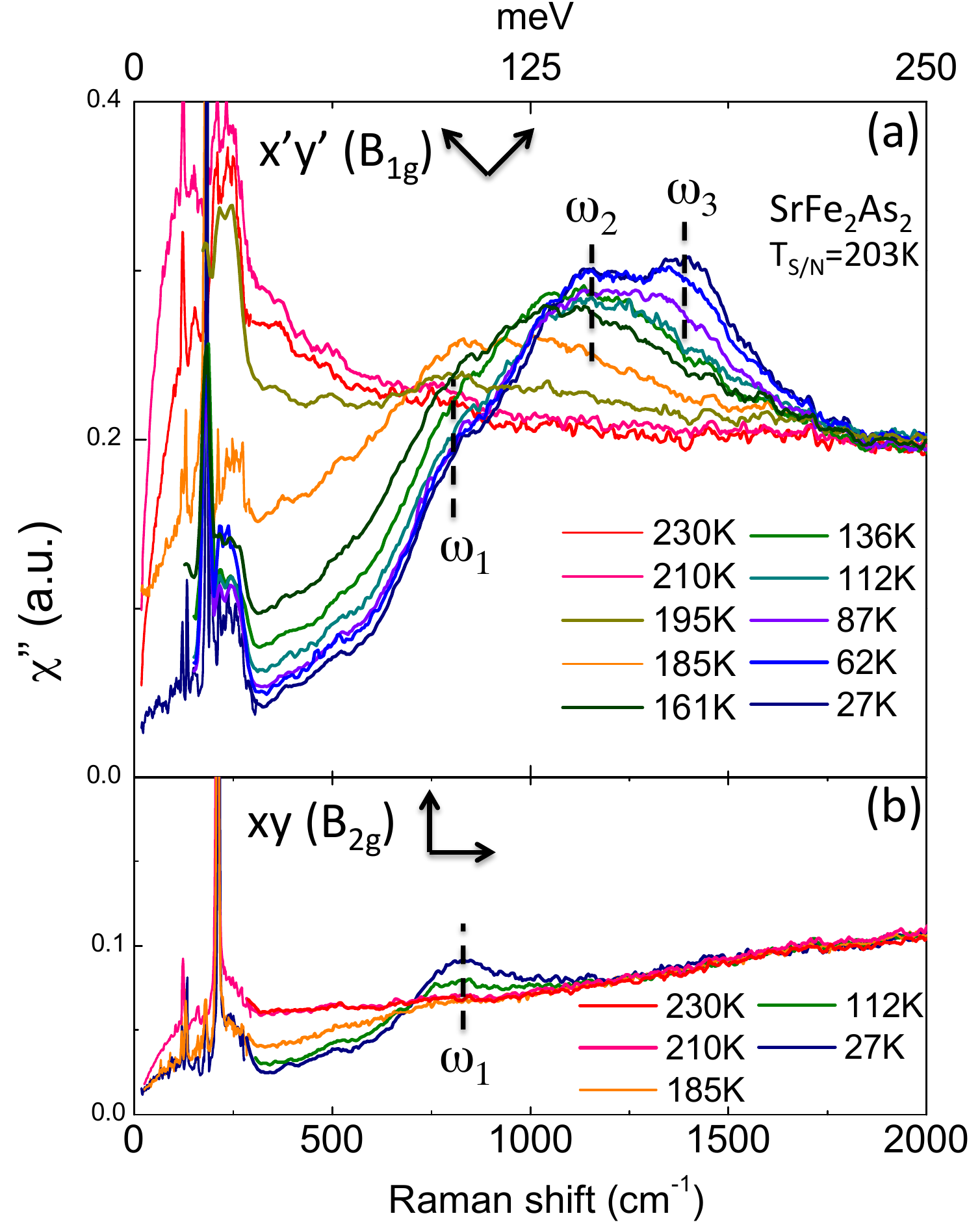}
\caption{Temperature dependence of the $x'y'$ (a) and $xy$ (b) Raman spectra in SrFe$_2$As$_2$ using 2.33~eV excitation energy.}
\label{fig3} 
\end{figure}

\begin{figure}
\centering \includegraphics[clip=,width=0.95\linewidth]{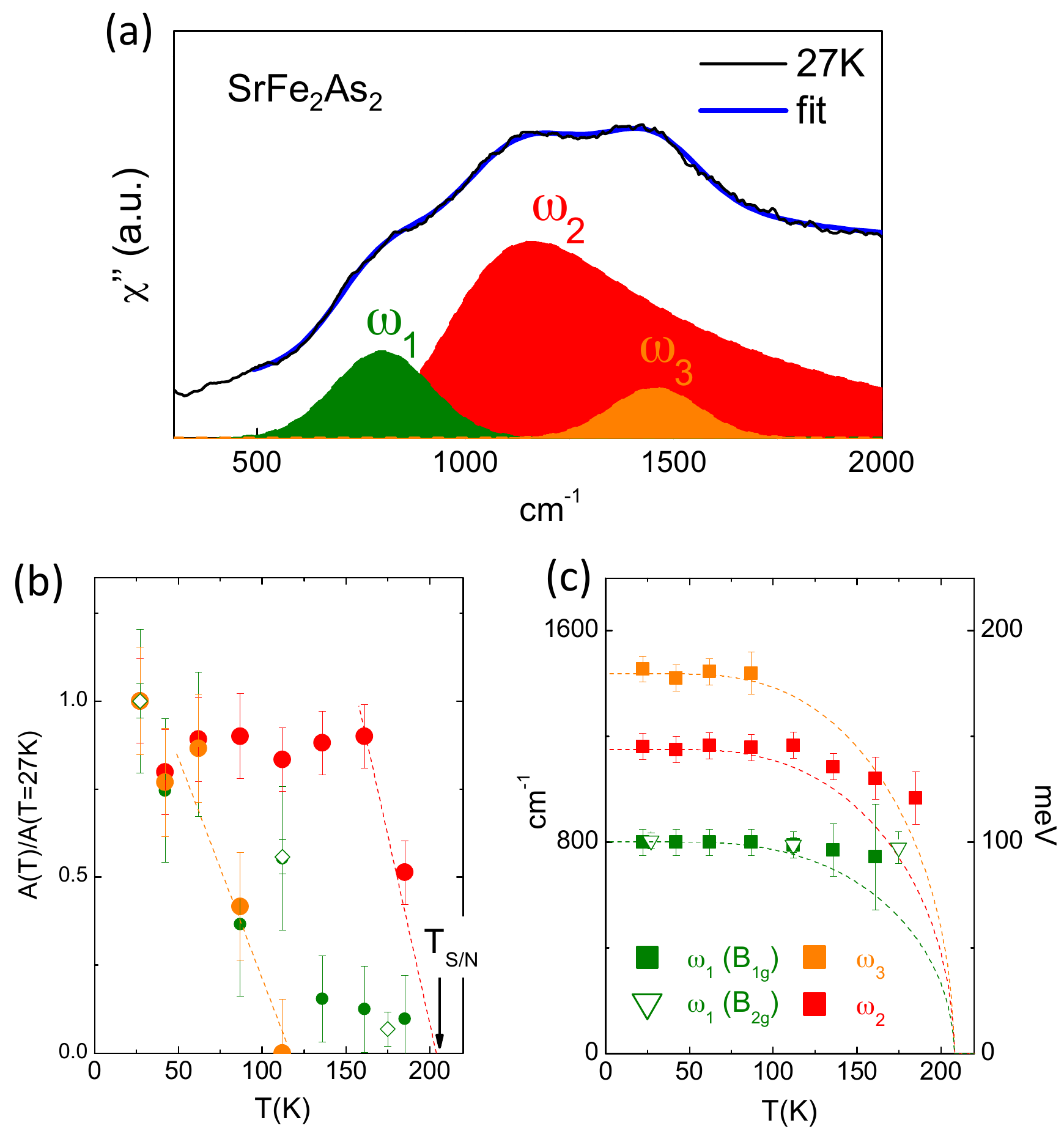}
\caption{(a) Decomposition of the $x'y'$ spectrum of SrFe$_2$As$_2$ at 27~K. Gaussian broadened asymmetric square-root singularities $\frac{1}{\omega\sqrt{\omega^2-\omega_0^2}}$, mimicking BCS-like coherence peaks \cite{Klein-Dierker} were used for each peak. A small linear background was added to account for the high energy continuum. The same analysis was performed for the $xy$ spectrum with only a single peak. (b) Evolution of the integrated area of each SDW gap peaks with temperature. (c) Evolution of the SDW gap peak energies with temperature. The dotted-lines show for comparison the mean-field BCS temperature dependences for each peak.}
\label{fig4} 
\end{figure}

\par
The detailed temperature dependence of the $x'y'$ and $xy$ spectra is shown in Fig. \ref{fig3}. It reveals a non-trivial evolution of the reconstructed electronic structure in the SDW state. As outlined above, $T_{N}$ marks the onset of the reshuffling of the Raman response, progressively transferring spectral weight from low to high energy in both symmetries. While most of the reshuffling occurs within less than 50~K below $T_{N}$, the different SDW peaks appear to emerge with markedly different temperature behaviors. This is particularly striking for the $\omega_3$ peak in the $x'y'$ spectrum which is clearly resolved only well below $T_{N}$, between 112~K and 87~K,  highlighting an unusual evolution of the electronic structure deep in the SDW phase. In order to gain a more quantitative evolution of the intensities of the SDW peaks below $T_{N}$, we have decomposed the $x'y'$ and $xy$ spectra using three and one asymmetric peaks respectively, along with a temperature dependent linear background. The results of the fits for the $x'y'$ spectrum at 27~K are shown in Fig. \ref{fig4}(a). The evolutions of the integrated area of the peaks and their energies are shown as a function of temperature in Fig. \ref{fig4}(b) and Fig. \ref{fig4}(c) respectively. The $\omega_2$ peak, which carries most of the spectral weight in the $x'y'$ spectrum, shows a rapid increase of integrated area right below T$_{N}$ before stabilizing below 160~K. Its energy softens upon approaching $T_{N}$, but the observed softening departs significantly from the mean-field weak-coupling BCS temperature dependence (shown in dotted-lines), which might be linked to the first order nature of the magnetic transition in Sr122 \cite{Li2009,Gillett}. By contrast the $\omega_1$ peak displays a more gradual increase of integrated area below $T_{N}$, while the $\omega_3$ peak area is sizable only below $\sim$100~K.  

\section{DISCUSSION AND COMPARISON WITH HALL MEASUREMENTS}

The above results appear to be broadly consistent with an itinerant picture of magnetism in 122 systems. The observation of multiple SDW gap peaks is expected in a multiband system where as many as 5 bands cross the Fermi level, resulting in multiple anti-crossing upon SDW induced band folding. The energies of the SDW gap peaks scale with $T_{N}$, and their symmetry dependence is consistent with SDW gaps forming around the $X$ and $Y$ points of the Brillouin zone, as expected from the locations of the hole and electron Fermi pockets in the paramagnetic state. The energy scales of the SDW gaps far exceed the BCS weak coupling ratio, but this is also the case for the few SDW systems where SDW gaps were studied so far, like chromium and 1D organics \cite{Fawcett,Vescoli}. The temperature dependence of the SDW peaks intensity however is intriguing as it clearly departs form the naive expectation that all gaps should open at once immediately below $T_{N}$.
\begin{figure}
\centering \includegraphics[clip,width=0.75\linewidth]{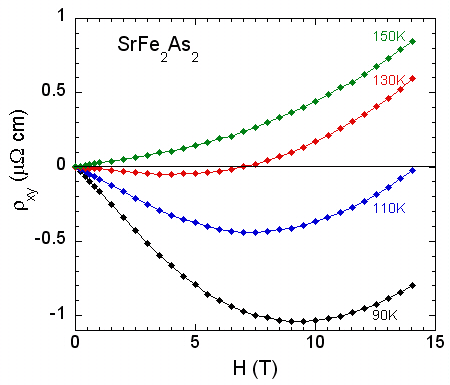}
\caption{Magnetic field dependence of the transverse (Hall) resistivity at 4 different temperatures below $T_N$ for SrFe$_2$As$_2$.}
\label{fig7} 
\end{figure}
\par
Given that neutron and thermodynamic measurements \cite{Yan,Zhao,Gillett} indicate only a single magnetic transition at $T_{N}$ in Sr122, the emergence of the $\omega_3$ peak well below $T_{N}$ seems surprising. As we argue below, it is most likely a fingerprint of a temperature dependent shift of the electronic structure deep in the SDW phase. Such unusual temperature effects on the electronic bands have recently been observed by ARPES experiments in the paramagnetic phase of Ba122 \cite{Dhaka,Brouet2013}. Part of the apparent band shift observed by ARPES could be a simple consequence of thermal population factors when the bottom or top of a band lies close to the Fermi energy \cite{Brouet2013}. However strongly temperature dependent band shifts may also result from the coupling between quasiparticles and interband spin fluctuations \cite{Benfatto}, or from a strong effect of lattice thermal contraction on the electronic structure \cite{Dhaka}.

\par
Because of the strongly reduced size of the Fermi pockets in the SDW state, these band shifts can result in drastic changes in the Fermi surface topology such as the progressive disappearance of hole or electron-like Fermi pockets upon cooling. Such changes in the topology of the Fermi surface should also affect other properties such as transport as was shown in the case of Co doping \cite{Rullier2009, Mun-Canfield-TEP}. In Fig. \ref{fig7} we show transverse (Hall) resistivity data $\rho_{xy}$ as a function of magnetic field at 4 different temperatures below $T_N$ for a Sr122 single crystal from the same batch. Due to the multiband character of the system, the transverse resistivity is highly non-linear above 5~T (see also \cite{Zentile2009}). Similar, albeit weaker, non-linear behavior was also found in undoped Ba122 \cite{Kuo-PRB2011}. Moreover we observe that this non-linearity is strongly enhanced when the quality of the samples is increased by annealing, as already reported for Ba122 \cite{Ishida2011}. Focusing on the low-field limit, the transverse resistivity indicate a clear change of the dominant type of carrier upon lowering temperature. While at 150~K the slope is positive, i.e. Hall coefficient $R_H>$0, indicating hole dominated transport, it changes sign at around 130~K and becomes strongly negative at lower temperature indicating electron dominated transport. The sign change of the low field Hall coefficient $R_H$ is for us the manifestation of a temperature induced change in the Fermi surface topology around 130~K, well below $T_N$. This might result in the gradual disappearance of a hole pocket occurring in a similar temperature range where the $\omega_3$ SDW gap peak emerges in the Raman data. It appears therefore that on the qualitative level, both Raman and transport data indicate a change in Fermi surface topology deep in the SDW phase. 
\par
It is interesting to note that a previous analysis of transport data on Co doped Ba122 found an unusual variation of carrier content with temperature in the SDW phase, which could also be ascribed to temperature dependent band shifts \cite{Rullier2009}. However in the case of Ba122, the Hall resistance remains electron-like at all temperatures and besides, the Raman spectra shows activated SDW gap peaks only immediately below $T_N$, indicating a less drastic effect of temperature on the Fermi surface topology than in the case of Sr122 \cite{Chauviere2011}. It seems that the more drastic effect of temperature on the Fermi surface topology in the case of Sr122 stems from the existence of a hole band whose top lies very close to the chemical potential in the reconstructed SDW electronic structure. Below we show via calculations of the SDW Raman response that the scenario of hole band sinking below the chemical potential upon cooling, as inferred from Hall data, is also qualitatively consistent with the evolution of the Raman spectra.

\section{THEORETICAL MODELING OF THE SDW RAMAN RESPONSE}

\subsection{Two-band model}
\begin{figure}
\centering \includegraphics[clip,width=0.99\linewidth]{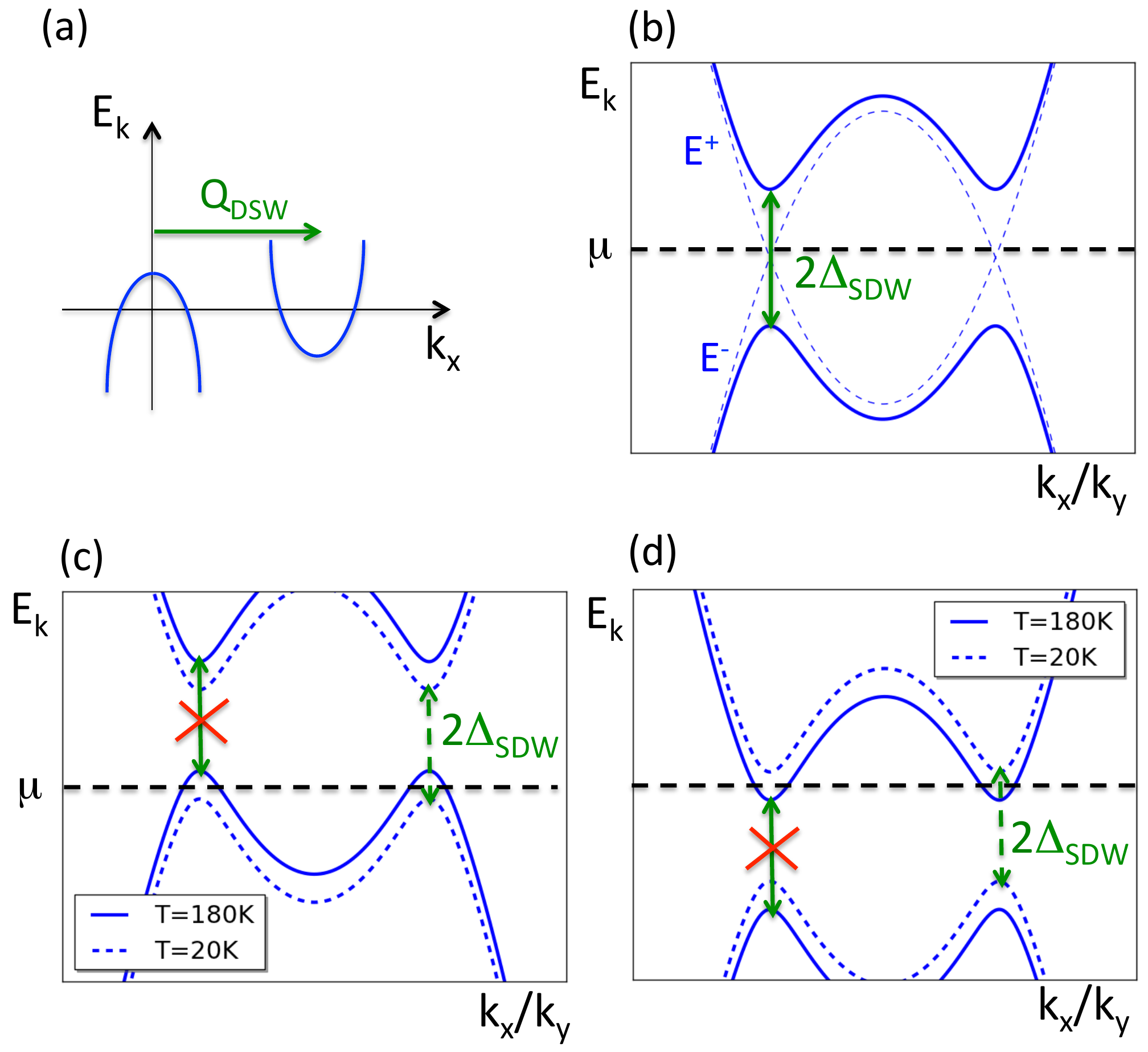}
\caption{ (a) Sketch of the dispersion of the electron and hole-like bands in the non-magnetic phase (b) Dispersion of the bands, $E_k^{\pm}$, in the SDW magnetic phase showing SDW gap opening. The folded non-magnetic bands are shown in dotted lines. (c)  Dispersion of the bands in the SDW phase with imperfect nesting leading to a residual hole-like Fermi pocket at 180~K, and quenched optical transitions across the SDW gap. Assuming a downward rigid band shift with lowering temperature, the hole pocket disappear and optical transitions across the SDW gap are activated. (d) Same scenario but with a residual electron pocket which disappear upon lowering temperature.}
\label{fig5a} 
\end{figure}

Being a two-particle probe, Raman scattering probes both the occupied and unoccupied part of the electronic dispersion. If a Fermi surface sheet disappears, interband optical transitions previously blocked by Pauli exclusion principle, become activated. As we will show below, the effect can be drastic at gapped anti-crossing points where the Raman coherence factors are enhanced, making Raman scattering extremely sensitive to changes in the topology of the small SDW reconstructed Fermi pockets as illustrated for different cases in Fig. \ref{fig5a}. Qualitatively, the enhancement of the Raman response can be understood by noting that electronic Raman scattering form factors (or vertex) are proportional to the electronic band curvature which is strongly enhanced at the folding induced anti-crossing points between hole-like and electron-like bands in reciprocal space \cite{Eiter2013}. 
\par
Here we present a simple two-band model showing the effect of changes in Fermi surface topology on the Raman response in the SDW phase. The starting point is a simplified two band model consisting of 2D circular hole and electron pockets with simple quadratic dispersions which are shifted by Q in momentum space (see Fig. \ref{fig5a}(a)):
 \begin{align}
 \epsilon_1(k)=\epsilon_1^0-\frac{\hbar^2 (k_x^2 +k_y^2)}{2m_h}   \\
 \epsilon_2(k+Q)=\epsilon_2^0+\frac{\hbar^2 (k_x^2 +k_y^2)}{2m_e} 
\end{align}
$\epsilon_1^0$ and $\epsilon_2^0$ are the top and bottom of the hole and electron bands respectively and $m_h$, $m_e$ their effective mass. A more realistic two-band model of the Fe SC would also include a finite ellipticity of the electron band. However since we will stay on the qualitative level, this will not affect our conclusions. The general expression for the imaginary part of the electronic Raman response is \cite{Dev-Hackl}:
\begin{multline}
(\chi^{\mu})''(\Omega)=\frac{2}{V}\sum_k\int^{\infty}_{-\infty}d\omega Tr \left[ \hat{\gamma}^{\mu}_k \hat{G_k}''(\omega)\hat{\gamma}^{\mu}_k \hat{G_k}''(\omega+\Omega)\right] \\
\times \left[f(\omega)-f(\omega+\Omega)\right]
\end{multline}

$\hat{G}''$ is the imaginary part of the Green's function $\hat{G}$ , $\hat{\gamma}^{\mu}_k$ the Raman form factor (or vertex) in the symmetry channel $\mu$ and $Tr$ is the trace. In the SDW phase, the Green's function  $\hat{G}$ can be written in 2x2 matrix form in Nambu space (see for example \cite{Fernandes} for a similar calculation of the optical conductivity):
\begin{multline}
\hat{G}(s,k,i\omega_n)=\frac{1}{(i\omega_n-E_k^+)(i\omega_n-E_k^-)} \\
\times\begin{pmatrix}
i\omega_n-\epsilon_2(k+Q) & -s\Delta_{SDW} \\
-s\Delta_{SDW} & i\omega_n-\epsilon_1(k)
\end{pmatrix}
\end{multline}
where $s$ denotes the sign of the spin, $i\omega_n$ is the Matsubara frequency, $\Delta_{SDW}$ is the SDW order parameter which couples electrons of band 1 at $k$ to electrons of band 2 at $k$+$Q$ and E$_k^{\pm}$ are the new quasiparticle dispersions:

\begin{multline}
E_k^{\pm}=\frac{1}{2}\left[\epsilon_1(k)+\epsilon_2(k+Q)\right] \\
\pm\sqrt{\Delta_{SDW}^2+\frac{1}{4}\left[\epsilon_1(k)-\epsilon_2(k+Q)\right]^2}
\end{multline}

In Nambu space, the Raman vertex $\hat{\gamma}_k^{\mu}$ is:
\begin{equation}
\hat{\gamma}_k^{\mu}=\begin{pmatrix}
\gamma_1(k) &  0 \\
0 & \gamma_2(k+Q)
\end{pmatrix}
\end{equation}

where $\gamma_1$ and $\gamma_2$ are the Raman vertices of individual bands in the symmetry $\mu$ centered at $k$ and $k$+$Q$ respectively. These vertices are proportional to the each individual band curvatures. Since our aim is not a realistic description of the symmetry dependent Raman response in iron-pnictides, but rather to illustrate the effect of band shifts near anti-crossing points in the SDW phase, we will make several assumptions on these vertices. First we will assume that these vertices do not depend on $k$ for each individual band. We will further assume that they have the same amplitude $\gamma_0$ but, due to the opposite band curvature of the two bands, have opposite signs: $\gamma_1(k)$=-$\gamma_2(k+Q)$ \cite{Vanyolos}. With these assumptions, and calling $G''_{ij}$ the matrix elements of $\hat{G}''$, we obtain the following expression for the imaginary part of the Raman response:

\begin{multline}
\chi''(\Omega)=\frac{2}{V}\sum_k\int^{\infty}_{-\infty}d\omega \gamma_0^2[G''_{11}(k,\omega)G''_{11}(k,\omega+\Omega) \\
+ G''_{22}(k,\omega)G''_{22}(k,\omega+\Omega) - 2G''_{12}(k,\omega)G''_{12}(k,\omega+\Omega)] \\
\times \left[f(\omega)-f(\omega+\Omega)\right]
\end{multline}

\begin{figure*}
\centering \includegraphics[clip,width=0.95\linewidth]{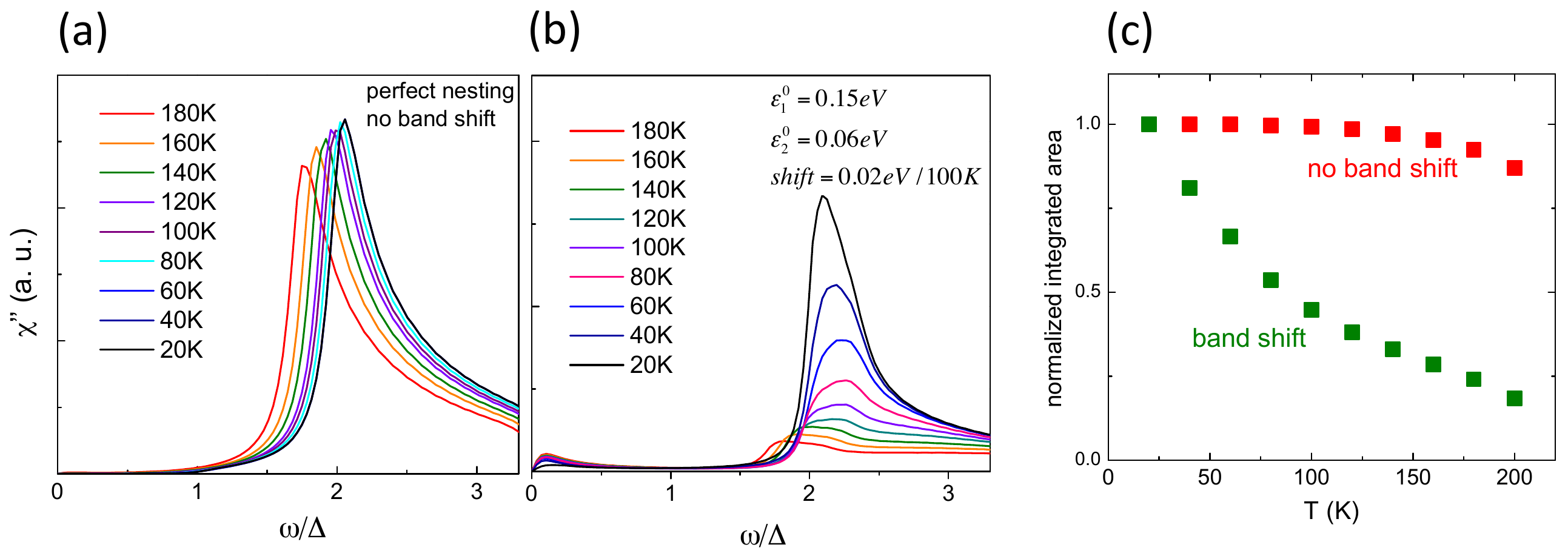}
\caption{(a) Calculated Raman spectra in the SDW phase as a function of temperature for perfect nesting $m_h$=m$_e$, $\epsilon_1^0$=-$\epsilon_2^0$ and for $\Delta_{SDW}$($T$=20~K)=70~meV. No band shift but a weak softening of the SDW gap energy close to $T_N$, consistent with the observed evolution of $\omega_2$ peak in the Raman experiment were assumed (see Fig. \ref{fig4}c). (b) Calculated temperature dependent Raman spectra for scenario (c) of Fig. \ref{fig5a} (the same spectra are obtained for scenario (d)). The following parameters have been used: $m_h$=$m_e$, $\epsilon_1^0$(180~K)=0.15~eV, $\epsilon_2^0$(180~K)=0.06~eV and $\Delta_{SDW}$(20~K)=0.07~eV. A downward rigid band shift of 0.02~eV/100~K was taken, mimicking qualitatively the motion of bands observed in ARPES experiments in the paramagnetic phase of Ba122 \cite{Brouet2013}. With these parameters the top of the hole-like band in the SDW state moves below the chemical potential below $\sim$80~K approximatively. (c) Area of the SDW gap Raman response as a function of temperature using the parameters of (b) (in green)). The same integrated area in a scenario where there is no band shift is shown for comparison in red. The area were normalized against their value at 20~K.}
\label{fig5b} 
\end{figure*}
The minus sign between the first two terms and the last term comes from the opposite curvatures of the two bands, electron-like and hole-like respectively, as expected for Fe SC. We note that for bands with equal curvature the sign would be positive. In that case one can show that there is a cancellation between the first two terms and the last term, yielding no contribution of SDW gap excitation to the Raman response \cite{Vanyolos}. There is analogy between these two opposite cases for the SDW response and coherence factors for type-I and type-II excitations in BCS theory. While the Raman response always involves type-I excitation processes in the BCS state, in the SDW phase however and depending on the relative sign of the band curvatures, the Raman response involves type-I (minus sign) or type-II (plus sign) processes.
 In order to compute the Raman response numerically, a small phenomenological scattering rate, $\Gamma=3.5~meV$, was included in the imaginary part of the self-energy in the Green's function.

\subsection{Effect of changes in surface topology on the SDW Raman response}
The case of perfect nesting between hole and electron pockets is illustrated in Fig. \ref{fig5a}(b). In this special case the chemical potential lies in the middle of the SDW gap. The calculated Raman is shown in Fig. \ref{fig5b}(a). It is dominated by a peak at 2$\Delta_{SDW}$, showing that the SDW Raman response predominantly comes from vertical transitions at the anti-crossing points in $k$ space due type-I excitations coherence factors. The integrated intensity of the peak is weakly temperature dependent in the SDW phase. It is controlled only by the energy of the SDW gap $\Delta_{SDW}$ which shows significant softening only very close to $T_N$. This temperature dependence is consistent with that of $\omega_2$ peak which shows a rapid increase of intensity below $T_N$ (see Fig. \ref{fig4}(b)). A very similar behavior of the SDW peak integrated intensity is also found in cases where the nesting is imperfect, as long as the band energies are temperature independent and/or they lie sufficiently far from the chemical potential. 
\par
By contrast, strong variations of the SDW gap peak intensity are found even well below $T_N$ when one of the bands sweeps across the chemical potential, because of temperature induced band shifts. Two scenarios, illustrated in Fig. \ref{fig5a}(c) and (d), can be envisioned for the activation of the SDW gap peak intensity upon decreasing temperature: either a hole-like band sinks below the chemical potential (c) or an electron-like band moves entirely above it (d). The corresponding calculated Raman spectra are identical in both cases and are shown in Fig. \ref{fig5b}(b). Initially quenched by Pauli blocking right below $T_N$, the SDW gap peak is weak and broad but gains considerable intensity and sharpens upon disappearance of the Fermi pockets. The effect is particularly strong because the SDW Raman response comes overwhelmingly from small regions of $k$ space, near the anti-crossing points. This makes Raman response much more sensitive to the changes in Fermi surface topology in the SDW state as compared to the paramagnetic state. The temperature evolution of the peak integrated intensity is displayed in Fig. \ref{fig5b}(c). In contrast with the essentially flat temperature dependence of the scenario with no band shift, the integrated intensity shows a strong and continuous increase between 180~K and 20~K.  We note that because of finite temperature effects which populate bands above the chemical potential, the temperature dependence of the integrated area does not show a sharp transition, but rather a crossover behavior. The continuous increase of intensity below $T_N$ is qualitatively consistent with the temperature dependence of the $\omega_3$ peak, whose intensity increases down to the lowest temperature measured (see Fig. \ref{fig4}(b)). We believe that its disappearance in the data above 100~K is the result of a combination of two effects: a spectral weight reduction due to band shifting effects (as illustrated in Fig. \ref{fig5b}(d)), and a strong broadening due to thermal population effects. While less spectacular, the behavior of the $\omega_1$ peak is also indicative of similar effects. A more realistic description of this band shifting effect on the Raman response is beyond the scope of this work, as it would require the knowledge of both the full reconstructed SDW band structure and the exact $k$-dependence of the multiband Raman vertex for Sr122. However it appears from the present calculations that the behavior of the $\omega_3$ peak is consistent with a band shifting induced Fermi pocket disappearance deep in the SDW state, in agreement with the Hall data of Fig. \ref{fig7}.
\par 
Our results thus indicate that the significant band shifting observed in previous ARPES measurements in the paramagnetic state persists in the magnetic state. Because of the strongly reduced size of the Fermi pockets in the SDW reconstructed state, the effect of band shifting can have more drastic consequences on the Fermi surface topology such as the Fermi pocket disappearance observed here. Being a probe of both occupied and unoccupied states, Raman data at a single doping cannot by themselves identify the nature of the vanishing pocket, hole or electron-like. Hall data however allow us to discriminate, and clearly favor the scenario of a vanishing hole pocket (as in Fig. \ref{fig5b}b). In addition the persistence of a two peak structure at $x=0.04$ with similar energy scales as $\omega_2$ and $\omega_3$ peaks observed at $x=0$, indicates the robustness of the $\omega_3$ peak upon electron doping. This is also consistent with the scenario of a hole, rather than electron, pocket lying close to the chemical potential in the SDW phase of Sr122. 
\par
Our Raman and Hall data on Sr122 are reminiscent of the case of Lifshitz transitions where changes in the Fermi surface topology such as "pocket-vanishing" or "neck-collapsing" are induced by doping or magnetic field at constant temperature \cite{Lifschitz,Imada,Vojta}. In the context of iron-arsenides superconductors, a doping induced Lifshitz transition was inferred from ARPES measurements in the SDW state of Co-Ba122, whereby a hole-like pocket was found to disappear upon electron doping \cite{Liu}. In the same system, Hall and thermopower measurements are also indicative of a change in Fermi surface topology in the SDW state upon increasing Co content \cite{Rullier2009, Wen-Mazin-Hall,Mun-Canfield-TEP}. 

\section{CONCLUSION}
Raman and Hall measurements reported here both indicate a change in the Fermi surface topology within the SDW phase of Sr122. The strong sensitivity of the SDW electronic structure with temperature, surmised in previous transport measurements on Ba122, is even more striking in the case of Sr122 and we have argued that it is most likely linked to the disappearance of a hole Fermi pocket upon cooling.
\par
It appears that the strong sensitivity of Fermi surface topology to various external parameters such as doping and temperature is a hallmark of Fe SC. An open question is how these changes may affect the superconducting ground state and in particular the pairing symmetry. There are already several theoretical investigations on possible changes of pairing symmetry in the strongly overdoped regime of K-Ba122, where electron Fermi pockets are expected to disappear \cite{Thomale2009,Thomale2011, Maiti2011,Maiti2011b, Kuroki2011, Maiti2013}. We suspect that the coexistence region between SC and SDW, with its reduced Fermi surface size, may offer an even more promising playground for the study of the interplay between Fermi surface topology and pairing symmetry \cite{Maiti-Fernandes2012}.

\section{ACKNOWLEDGMENTS}
The authors thank L. Bascones, V. Brouet, I. Paul and B. Valenzuela for helpful discussions. We thank B. L\'eridon for her help with SQUID magnetometer measurements.This work was supported by Agence Nationale de la Recherche through ANR Grant PNICTIDES and by a SESAME grant from R\'egion Ile-de-France.

\bibliography{ref}

\end{document}